\documentclass[%
preprint,
 amsmath,amssymb,
 aps,
prf,
]{revtex4-2}

\usepackage{graphicx}
\usepackage{dcolumn}
\usepackage{bm}
\usepackage{comment}
\usepackage{hyperref}
\hypersetup{hidelinks}
\usepackage{subcaption}
\usepackage[dvipsnames]{xcolor}

\begin{document}

\preprint{APS/123-QED}

\title{An unusual bifurcation scenario in a stably stratified, valley-shaped enclosure heated from below}

\author{Patrick J. Stofanak}
\author{Cheng-Nian Xiao}
\author{Inanc Senocak}%
 \email{senocak@pitt.edu}
\affiliation{%
 Department of Mechanical Engineering and Materials Science,\\
 University of Pittsburgh, Pittsburgh, PA 15261, USA
}%


\date{\today}

\begin{abstract}
We delineate the structure of steady laminar flows within a stably stratified, valley-shaped triangular cavity heated from below through linear stability analysis and Navier-Stokes simulations. 
Our formulation follows Prandtl's idealization of mountain and valley flows. We derive an exact solution to the quiescent conduction state, which bifurcates to two pairs of flow patterns with symmetric and asymmetric circulations in the valley-shaped enclosure.
We characterize the flow  via the stratification perturbation parameter, $\Pi_s$, which is a measure of the strength of the surface heat flux relative to the strength of the background stable stratification, while keeping other dimensionless parameters fixed. 
At very low $\Pi_s$ values, the pure conduction state remains stable.  
Beyond a threshold value of $\Pi_s$, a unique bifurcation event transpires, giving rise to two discernible types of eigenmodes. One type is marked by a single dominant circulation positioned at the valley's center, while the other one exhibits dual circulations of equal strength within the same central valley region. Notably, the critical value associated with the central-circulation eigenmode is marginally lower than that of the dual-circulation eigenmode. Through three-dimensional Navier-Stokes simulations, we confirm that the central-circulation eigenmode generates a pair of asymmetric steady states, whereas the dual-circulation eigenmode leads to distinct upslope and downslope symmetric convection patterns. 
Linear stability analysis and three-dimensional Navier-Stokes simulations jointly confirm the linear instability of the two dual-circulation states, indicating that the slightest asymmetric perturbation will induce a transition towards the asymmetric steady state, which is the primary state for the parameter values considered here.
Overall, our investigation reveals that, for a given set of dimensionless parameters, the Navier-Stokes equations admit at least five possible steady-state solutions. Two of these solutions, namely the quiescent, pure conduction state and the counter-intuitive symmetric downslope state, have previously been overlooked in stably stratified valley-shaped enclosures heated from below.
Taken together, these five flow solutions reveal an intriguing bifurcation structure, encompassing both a perfect pitchfork bifurcation and a nested bifurcation that gives rise to two distinct states. The inner bifurcation, while resembling a pitchfork in some respects, does not break any symmetry of the valley due to the lack of any possible horizontal axis of symmetry. The categorization of this inner bifurcation remains an unresolved matter, as it does not conform to any established descriptions of canonical bifurcations.


\end{abstract}

\keywords{stratified flows, pitchfork bifurcation, anabatic slope flows}
\maketitle


\section{\label{sec:level1} Introduction} 

During an atmospheric evening transition in regions of complex terrain, surface cooling induces downslope, or katabatic winds. These katabatic winds contribute to the development of a stably stratified cold pool in valleys, which persists throughout the night. During the morning transition, surface heating prompts upslope, or anabatic flows, opposing the stably stratified cold pool's presence and eventually causing its dissolution. Notably, numerical weather prediction models often encounter challenges when dealing with stably stratified flows in complex terrain \citep{holtslag2013stable}, as well as the transitions between nocturnal stably stratified periods and daytime intervals \citep{angevine2020transition}. These limitations can adversely impact the accuracy of predictions regarding morning fog formation and pollutant dispersion \citep{boutle2018aerosol,salmond2005review}. Consequently, our objective is to enhance the comprehension of stably stratified flows experiencing surface heating within an idealized valley setting.

In this study, we adopt Prandtl's idealization for mountain and valley flows that assumes a constant ambient or background stratification independent of the thermal forcing at the surface \cite{prandtl1942, prandtl1953essentials}. To this end, the presence of an ambient stable stratification sets our problem apart from Rayleigh-B\'enard type convective flows because of the expanded set of dimensionless numbers governing the flow problem. Our choice of a V-shaped enclosure represents an idealized geometry for valley flows, facilitating the examination of fundamental instabilities and flow patterns. Additionally, this selection establishes connections with prior experimental research \citep{princevac2008morning, wang2021experimental} and lays the groundwork for forthcoming experiments concerning flow regimes and instabilities.


Thermal convection in attic-shaped (an inverted valley geometry) triangular cavities with isothermal conditions on sloped walls without any stratification effects have been studied \citep{saha2011review}. In such configurations, a symmetric convection pattern prevails at low Grashof numbers, and a subcritical pitchfork bifurcation occurs at larger parameter values leading to a steady asymmetric circulation state \citep{ridouane2006formation,omri2007numerical}, which has been shown to agree with smoke flow visualizations in an attic-shaped triangular cavity with a finite depth \citep{holtzman2000laminar}.

In comparison to  attic-shaped cavities, there has been relatively less attention on convection in V-shaped enclosures with ambient stratification.  \citet{princevac2008morning} conducted experiments with stratified saline water in a V-shaped tank heated with a constant heat flux on bottom walls to represent morning transition in valleys in a laboratory setting. They introduced the dimensionless breakup parameter $B$, along with the slope angle of the valley walls, to characterize flow patterns that form along the sloping walls. 
Motivated by the experiments of \citeauthor{princevac2008morning}, \citet{bhowmick2018natural} used two-dimensional (2D) Navier-Stokes (N-S) simulations to investigate flow dynamics in triangular V-shaped cavities heated from below with an initially stratified fluid and adiabatic conditions on the top boundary. Their 2D simulations produced symmetric circulation patterns within the valley-shaped cavity. \citet{bhowmick2019transition,bhowmick2022chaotic} also performed 2D N-S simulations for valley-shaped cavities heated from below and cooled from the top boundary without any stratification effects. A common dynamics that was observed in these 2D N-S simulations as a function of increasing Rayleigh number is the establishment of a steady symmetric circulation transitioning to a steady-state asymmetric circulation through a pitchfork bifurcation, which is proceeded by the emergence of a periodic state through a Hopf bifurcation. 


Our present investigation establishes additional connections with aforementioned studies that involve convection in triangular cavities and other widely recognized convection phenomena. Particularly pertinent to our study is the presence of multiple feasible steady-state solutions in a convection problem with an ambient stable stratification. Multiple steady-state solutions in convection problems were demonstrated by \citet{gelfgat1999stability} in the context of Rayleigh-B\'{e}nard convection, and likewise for other confined, convective flows \citep{erenburg2003multiple}. \citeauthor{venturi2010stochastic} \cite{venturi2010stochastic} illustrate that in Rayleigh-B\'{e}nard convection, the single-roll convection pattern is often favored over the two-roll state due to its heightened heat transport efficiency. This tendency is similarly observable in geophysical models of ocean dynamics. \citeauthor{marotzke1988instability} \cite{marotzke1988instability} employ an idealized meridional-plane model to investigate ocean thermohaline circulation, uncovering the existence of multiple equilibrium states, including an unstable symmetric state that transitions into an asymmetric state upon minor perturbation. Correspondingly, \citeauthor{bryan1986high} \cite{bryan1986high} scrutinizes the same configuration through a three-dimensional general circulation model of the ocean, yielding analogous findings.


Several dimensionless parameters have found use in prior studies for characterizing stratified flows with surface heating, as well as stratified flows in cavities. In a series of papers, \citet{xiao2019stability, xiao2020stability} established the stratification perturbation parameter $\Pi_s$ in the context of Prandtl's idealization of mountain and valley flows. This parameter signifies the ratio between the imposed surface heat flux and the stabilizing background stratification, and it serves as a basis for assessing the dynamical stability of the Prandtl model concerning katabatic and anabatic slope flows. In the context of stratified flows within cavities, \citet{yalim2018vertically} employed the buoyancy number \textemdash a measure of the ratio between viscous and buoyancy timescales \textemdash to examine the instability of a stably-stratified fluid within a square cavity undergoing vertical oscillations. Similarly, \citet{grayer2020dynamics}  utilized this parameter to explore the dynamics of stably stratified flow in a tilted square cavity with differentially heated side walls. Furthermore, the experimental work conducted by \citet{princevac2008morning} in a stratified water tank heated from below introduced the so-called break-up parameter $B$ to track the disintegration of stratification within an initially stably stratified V-shaped valley. In the next section, we show that the break-up parameter is expressible as a combination of the stratification perturbation parameter, the buoyancy number, and the Prandtl number.

The current study examines instabilities and steady-state convection patterns within a stably stratified valley-shaped enclosure heated from below. We investigate a set of conditions that have not been previously explored using linear stability analysis and three-dimensional (3D) simulations of the Navier-Stokes equations. Specifically, we extend the dimensionless parameter space for the studied configuration, scrutinizing transitions between various potential flow states within a multi-stable arrangement.
We apply a constant positive heat flux to both bottom walls of the valley-shaped enclosure, which parallels the experimental setup outlined in \citet{princevac2008morning}. Simultaneously, we enforce a fixed temperature at the upper boundary, thereby establishing conditions conducive to a pure conduction state. Utilizing the stationary, pure conduction steady state as our starting point in these investigations holds a distinct advantage. In such cases, linear stability and nonlinear energy stability are in agreement \citep{shir1968convective}. This alignment allows us to comprehensively capture the exact bifurcation of the flow at the first critical stability threshold through linear stability analysis\textemdash an outcome corroborated by our validation through 3D Navier-Stokes simulations. Additionally, our approach encompasses a constant background stable stratification, unaffected by the thermal forcing at the surface. This approach aligns with Prandtl's idealization of mountain and valley flows \citep{prandtl1942, xiao2019stability}, and thus the current study represents an extension of Prandtl's model to anabatic flows in idealized valleys.
Although similarities can be drawn between our current arrangement and the dynamics observed in comparable scenarios, such as Rayleigh-B\'{e}nard convection, the distinctive factor in our present study is the introduction of an extra dimensionless parameter due the presence of a constant stable stratification of the ambient environment, which adds a forcing term to the buoyancy equation, which is absent in typical Rayleigh-B\'{e}nard configurations. 
Notably, \citet{xiao2022impact} demonstrated that this ambient stratification plays a crucial role in modeling stably stratified flows, even over flat surfaces.

\section{Technical formulation}

\begin{figure}
    \centering
    \includegraphics[width=0.66\textwidth]{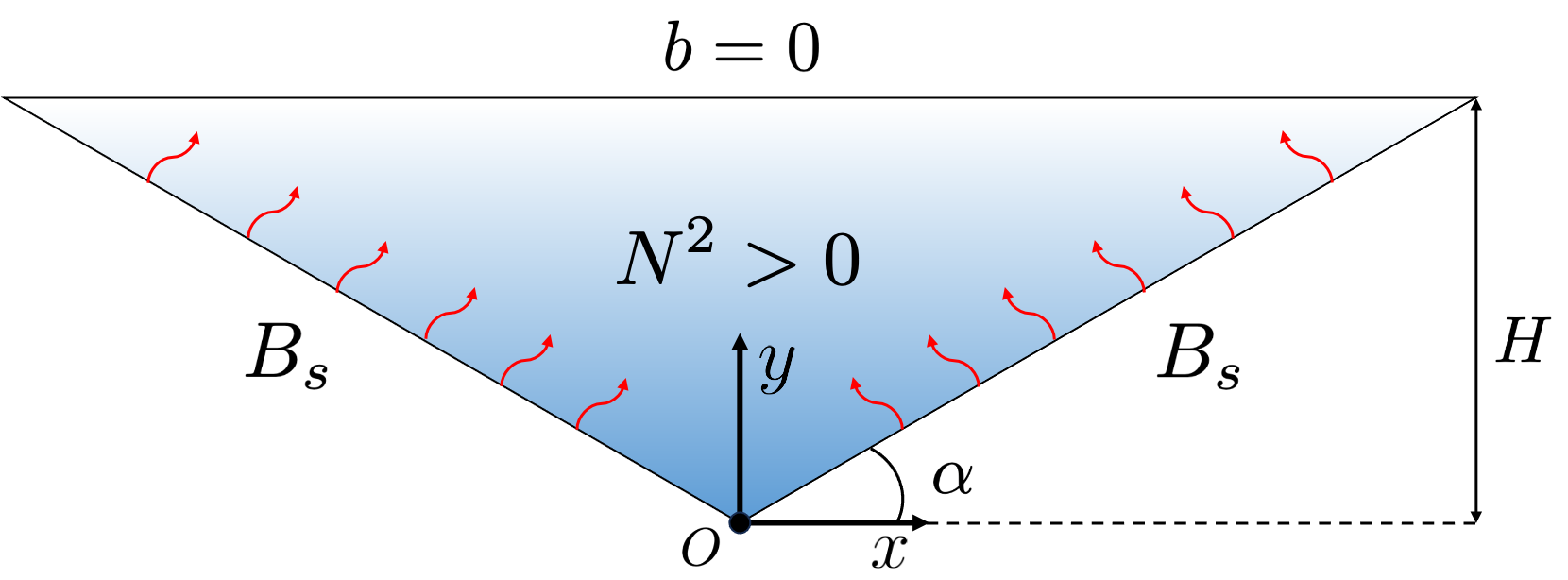}\hfill
    \caption{Schematic of the computational domain with key external parameters. The $x$ and $y$-axes are shown at the origin with the positive $z$-axis being out of the page. All simulations and visualisations adopt these axes.
    }
    \label{fig:ValleySchematic}
\end{figure}

A schematic of the computational domain is shown in Figure \ref{fig:ValleySchematic}, where $H$ is the height of the domain and $\alpha$ is the slope angle of the walls of the valley-shaped enclosure relative to the $x$ axis.  
The enclosure lies in the $x-y$ plane, with the homogeneous $z$ direction into the page. 
Thus $u$ represents the horizontal velocity in the $x$ direction, $v$ represents the vertical velocity in the $y$ direction, and $w$ represents the spanwise velocity in $z$ direction. 
We perform linear stability analysis (LSA) in only the two-dimensional (2D) valley geometry, and thus we only present 2D instabilities resulting from our LSA. However, our Navier-Stokes simulations are conducted with all three dimensions resolved to ensure that the observed instabilities are two dimensional.

The buoyancy is related to the potential temperature $\Theta$ by $b = g \left( \Theta - \Theta_e \right) / \Theta_r$, where  $\Theta_e$ is the potential temperature of the ambient environment, and $\Theta_r$ is a reference potential temperature. A constant background stratification is imposed through the buoyancy frequency, or Brunt-V\"ais\"al\"a frequency, given by $N = \sqrt{ \left(g / \Theta_r\right) \partial \Theta_e / \partial y}$ in the buoyancy transport equation (see the last term in equation \ref{eq:buoy}). Thus $b$ represents a scaled perturbation of the full potential temperature $\Theta$ from the linear background stratification defined by $\Theta_e$, manifested in the constant $N$.

The continuity, momentum, and buoyancy equations, with the Boussinesq approximation, can be written as follows:
\begin{equation} \label{eq:cont}
    \nabla \cdot \mathbf{u} = 0,
\end{equation}
\begin{equation} \label{eq:mom}
    \frac{\partial \mathbf{u}}{\partial t} + \mathbf{u} \cdot \nabla \mathbf{u} = - \nabla p + \nu \nabla^2 \mathbf{u} + b \mathbf{g},  
\end{equation}
\begin{equation} \label{eq:buoy}
    \frac{\partial b}{\partial t} + \mathbf{u} \cdot \nabla b = \beta \nabla^2 b - N^2 \mathbf{g} \cdot \mathbf{u},  
\end{equation}
where $p$ is the specific pressure including a constant reference density $\rho_0$, $\nu$ is the kinematic viscosity, $\beta$ is the thermal diffusivity, and $\mathbf{g}$ represents the effective gravity vector, $\mathbf{g} = [0, 1, 0]$, acting only in the $y$ direction. 
Here, the last term of equation (\ref{eq:buoy}) should be noted, as this additional term in the buoyancy equation arises due to the constant background stratification defined by $N$. This term arises from the formulation by \citeauthor{prandtl1942} \cite{prandtl1942} and separates the current analysis from simple Rayleigh-B\'{e}nard convection in a triangular cavity, as pointed out by Prandtl. 

The boundary conditions for buoyancy include a constant, positive buoyancy flux on the two bottom walls, defined as $B_s = \beta \partial b / \partial n$, where $n$ is the direction normal to the sloping bottom boundaries, and where a positive $\partial b/ \partial n$ refers to heating of the fluid. On the top boundary, a constant $b = 0$ is imposed. For velocity, a no-slip condition is imposed on the two bottom walls, and a free-slip condition is imposed on the top boundary.
These boundary conditions are chosen to parallel the experimental study of \citeauthor{princevac2008morning} \cite{princevac2008morning} which consists of water in a V-shaped tank with a free surface on top. However, unlike the experiments which consist of high Reynolds number flows, we focus on the dynamics near the stability threshold, meaning we only investigate laminar flows in the absence of an ambient wind.  


Under these conditions, the following exact solution with zero velocity for $p$ and $b$ to Equations \ref{eq:mom} and \ref{eq:buoy} can be derived
\begin{equation} \label{eq:exact_buoy}
p(y) = \frac{- B_s}{2 \beta \cos \alpha} \left( y - H \right)^2, \qquad
b(y) = \frac{B_s}{\beta \cos \alpha} \left( H - y \right).
\end{equation}
This motionless steady state within the heated valley is only possible  due to the constant buoyancy imposed at the top boundary combined with the constant heat flux at the sloped surfaces, which admits  a linear buoyancy and quadratic pressure profile as a solution. 
Equation (\ref{eq:exact_buoy}) represents the motionless, pure conduction state in the enclosure and will be used as a base flow for our linear stability analysis.
For 3D nonlinear simulations, all variables are periodic in the homogeneous $z$ direction with a length of twice the height of the valley-shaped enclosure.

We use the following scales to normalize dimensional quantities:
\begin{equation} \label{eq:scales}
    l_0 = H, \quad u_0 = \sqrt{\frac{B_s}{N}}, \quad b_0 = \frac{B_s H}{\beta \cos \alpha}, \quad p_0 = \frac{B_s H^2}{\beta \cos \alpha} , \quad t_0 = H \sqrt{\frac{N}{B_s}}, 
\end{equation}
where the length scale, $l_0$, is defined as the height of the valley geometry, and the timescale is the convection timescale, $t_0 = l_0/u_0$. An alternative length scale could be defined as half of the valley width, $l_0 = H/\tan \alpha$, and the stratification timescale can be defined as $t_0 = 1/N$. The height was chosen as the length scale to more conveniently non-dimensionalize the base flow, and the convective timescale is used throughout the rest of the present study as it better describes the timescale of the evolution of the present instabilities. 
The vorticity is defined as the curl of the velocity, $\mathbf{\omega} = \nabla \times \mathbf{u}$, but due to the purely 2D nature of the instabilities investigated here, we refer only to the vorticity in the $z$ direction, $\omega_z$, with a corresponding scale $\omega_0 = u_0 / l_0$.

Using the physical scales defined in equation (\ref{eq:scales}), we non-dimensionalize the governing equations (\ref{eq:cont}-\ref{eq:buoy}) and obtain the following:
\begin{equation} \label{eq:cont_nondim}
    \nabla \cdot \mathbf{u} = 0,
\end{equation}
\begin{equation} \label{eq:mom_nondim}
    \frac{\partial \mathbf{u}}{\partial t} + \mathbf{u} \cdot \nabla \mathbf{u} = - \frac{\Pi_h}{\cos \alpha} \nabla p + \frac{Pr}{\sqrt{\Pi_s \Pi_h}} \nabla^2 \mathbf{u} + \frac{\Pi_h}{\cos \alpha} b \mathbf{g}, 
\end{equation}
\begin{equation} \label{eq:buoy_nondim}
    \frac{\partial b}{\partial t} + \mathbf{u} \cdot \nabla b = \frac{1}{\sqrt{\Pi_s \Pi_h}} \nabla^2 b - \frac{\cos \alpha}{\Pi_s} \mathbf{g} \cdot \mathbf{u}. 
\end{equation}
Similarly, the analytical solution for the pure conduction state simplifies to 
\begin{equation} \label{eq:exact_buoy_norm}
p(y) = -\frac{1}{2} \left( y - 1 \right)^2, \qquad
b(y) = 1 - y ,
\end{equation}
where all variables are now in their non-dimensional form. From equations (\ref{eq:cont_nondim}-\ref{eq:buoy_nondim}), we see that flow in a valley-shaped enclosure under stable stratification is controlled by four dimensionless parameters, defined as,
\begin{equation} \label{eq:dim_par}
    \Pi_s = \frac{B_s}{\beta N^2}, \quad \Pi_h = \frac{N H^2}{\beta}, \quad Pr = \frac{\nu}{\beta}, \quad \alpha,
\end{equation}
where $Pr$ is the Prandtl number, and $\alpha$ is the slope angle. The dimensionless stratification perturbation parameter, $\Pi_s$, first introduced to characterize the stability of Prandtl slope flow \citep{xiao2019stability} is key to the present investigation. $\Pi_s$ is a measure of the strength of the imposed surface buoyancy gradient relative to the stabilizing background stratification. In Prandtl slope flows, the higher the $\Pi_s$ the more dynamically unstable and turbulent the flow becomes \cite{xiao2020stability}.
The dimensionless parameter, $\Pi_h$, is a new addition and related to the buoyancy number $R_N$ \citep{yalim2018vertically,grayer2020dynamics} by $\Pi_h = Pr R_N$, and represents the ratio between the thermal diffusion and buoyancy time scales.

We note that the set of dimensionless parameters given in equation (\ref{eq:dim_par}) is larger than the set adopted in previous studies of stratified flows in valley-shaped enclosures. For example, \citet{princevac2008morning} introduce the dimensionless breakup parameter $B = N^3 H^2 / B_s$, along with $Pr$ and $\alpha$, whereas the Rayleigh number was used in \citet{bhowmick2018natural}. In light of the expanded parameter space given in equation (\ref{eq:dim_par}), we observe that $B$ is a combination of two independent dimensionless parameters $B = \Pi_h / \Pi_s$. In this way, we can say that four dimensionless parameters are needed to fully describe the flow dynamics. 

\subsection{Linear stability analysis}

We linearize equations (\ref{eq:cont}-\ref{eq:buoy}) around an arbitrary base flow defined by $(U_i, \bar{p}, \bar{b})$, and assume disturbances take the form of waves given by
\begin{equation}
    \mathbf{\hat{q}}(x, y, t) = \left[ \hat{u}(x,y), \hat{v}(x,y), \hat{w}(x,y), \hat{p}(x,y), \hat{b}(x,y) \right] \exp \left( \omega t \right),
\end{equation}
where $\mathbf{\hat{q}}$ represents the vector of 2D disturbance quantities, and  $\omega$ represents the temporal growth rate. Substitution of the above disturbance quantities into the linearized Navier-Stokes equations leads to the following generalized eigenvalue problem
\begin{equation} \label{eq:eig_prob}
    \mathbf{A} \mathbf{\hat{q}} (x, y) = \omega \mathbf{B} \mathbf{\hat{q}} (x, y).
\end{equation}
By solving the eigenvalue problem, we can determine the global linear stability behavior of the given base flow for a 2D valley-shaped enclosure. The real part of the growth rate, $\mathrm{Re}(\omega)$, indicates whether an infinitesimal disturbance will exponentially grow, when $\mathrm{Re}(\omega) > 0$, or decay, when $\mathrm{Re}(\omega) < 0$, while the imaginary part, $\mathrm{Im}(\omega)$, indicates the temporal frequency of the resulting mode of instability. All simulations were carried out using  the spectral/hp element code N\textsc{ektar}++ \citep{cantwell2015nektar++,moxey2020nektar++}, which has been widely validated for a variety of flow problems. All simulations are run with a finite element discretization of 31 elements along the bottom walls with a polynomial order of four within each element. The eigenvalue problem is solved using the modified Arnoldi method available in Nektar++, and it is found that refinement in our spectral element mesh results in very little to no change in the eigenvalue (0.1\% difference near the critical value), which provides confidence in our results. 

Numerical integration of the full 3D Navier-Stokes equations was performed with Nektar++ to produce steady-state profiles, as well as to obtain and validate secondary states arising from the primary instabilities. For 3D simulations, we resolve the dimensionless spanwise extent of the valley ($L_z=2$) with 32 Fourier modes. Navier-Stokes simulations are performed with both initially 2D disturbances, defined by the eigenmodes obtained from LSA, as well as 3D random disturbances. The inclusion of the third direction in the Navier-Stokes simulations is to ensure the 2D dynamics observed are not based on a lack of consideration of the third direction.


\section{Results}

\subsection{Primary linear stability analysis}

We first perform two-dimensional (2D) linear stability analysis (LSA) of the valley-shaped enclosure with a zero velocity base flow with the pressure and buoyancy profiles given in equation (\ref{eq:exact_buoy}).
In all the simulations and analyses, slope angle $\alpha$ and $\Pi_h$ are fixed at $30^\circ$ and $1500$, respectively. Prandtl number $Pr$ is set to $7.0$ to parallel the experimental study of \citet{princevac2008morning}. We only vary $\Pi_s$ throughout the study.

\begin{figure*}
    \centering
    \includegraphics[width=0.96\textwidth]{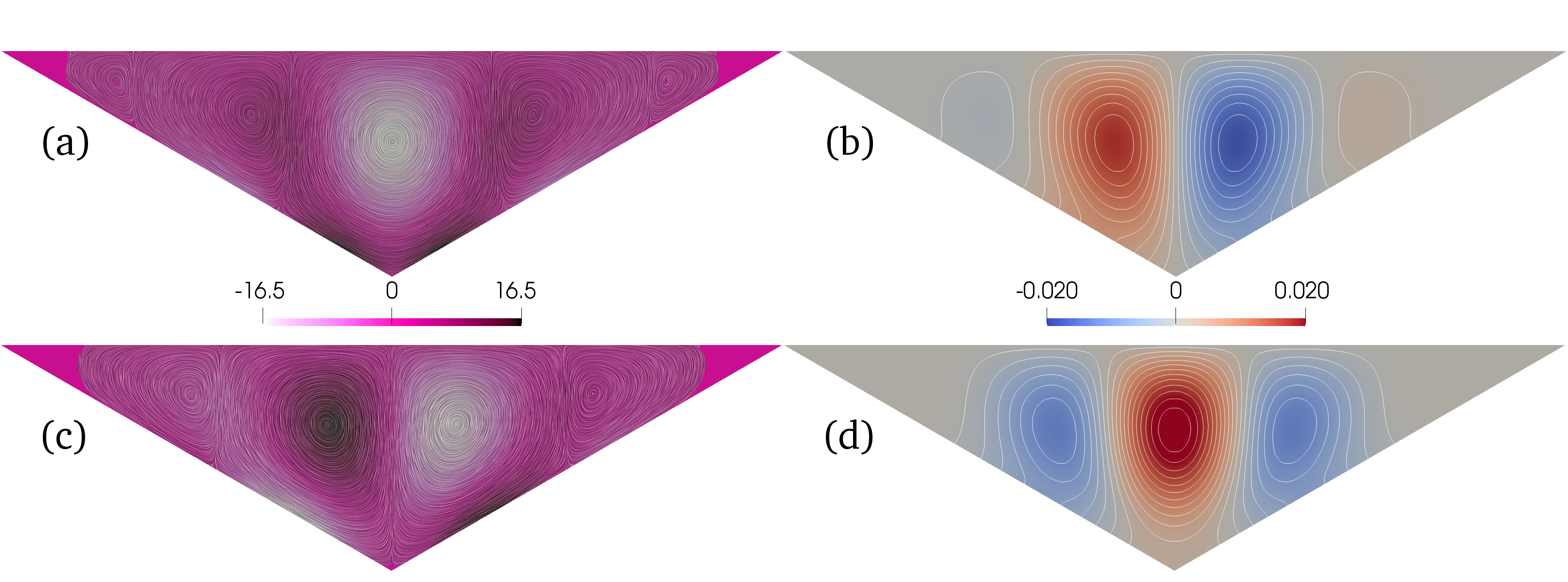}\hfill
    \caption{Visualization of eigenmodes resulting from linear stability analysis of analytical base flow for $\Pi_s = 0.9$, showing (a) vorticity and (b) buoyancy of the central-circulation eigenmode, and (c) vorticity and (d) buoyancy of the dual-circulation eigenmode. 
    Line integral convolution (SurfaceLIC) is used to visualize vector fields, and buoyancy profiles are shown with isotherms.
    }
    \label{fig:eigenmodes}
\end{figure*}

For small values of $\Pi_s$, meaning the perturbation caused by the surface buoyancy flux is small compared to the stabilizing background stratification, the quiescent, pure conduction state is stable. 
As we increase $\Pi_s$, this base  state becomes linearly unstable, and LSA reveals two eigenmodes at critical $\Pi_s$ values, shown in Figure \ref{fig:eigenmodes}. The imaginary part of the eigenvalue of both modes is zero, indicating they are non-oscillatory. 

The vorticity field and streamlines for the first eigenmode, displayed in Figure \ref{fig:eigenmodes}(a),  shows that it consists of one large circulation in the center  alongside smaller, counter-rotating corner vortices. 
The  buoyancy profile, shown in Figure \ref{fig:eigenmodes}(b), depicts how the central circulation advects heat away from the hot bottom walls and how the colder fluid near the top wall recirculates down towards the surface. 
For these reasons, we refer to this eigenmode as the central-circulation eigenmode. 
An examination of the velocity and buoyancy field of the central-circulation eigenmode reveals that it satisfies the symmetry of reflection about the $y$ axis for the linearized N-S equations. Specifically, this can be defined by the transformation $\mathcal{C}_1$:
\begin{equation}
    \mathcal{C}_1 : \left[u, v, b, p \right] \left(x, y \right) \mapsto \left[u, -v, -b, -p \right] \left(-x, y \right).
\end{equation}
The central-circulation eigenmode is invariant under this transformation, and thus we refer to it as symmetric with respect to this transformation, or $\mathcal{C}_1$-symmetric in short. However, we note that while the linearized N-S equations are invariant under this transformation, the full N-S equations 
are not invariant due to the nonlinear advection terms, and therefore this symmetry must be broken in the nonlinear evolution of the eigenmode.

The second eigenmode, shown in Figures \ref{fig:eigenmodes}(c) and \ref{fig:eigenmodes}(d),  consists of two identical, counter-rotating main circulations on each side of the  valley center. 
The buoyancy profile, Figure \ref{fig:eigenmodes}(d), shows that the direction of the two dominant circulations is downslope, which can be seen through the high temperature in the center of the valley. 
This suggests that the eigenmode exists in both upslope and downslope configurations. This is further supported by the fact that we find no additional unstable eigenvalues from our LSA. While the emergence of two possible states from a given eigenmode is not surprising, we note here that the downslope configuration represents flow traveling down the heated bottom walls, which is a counter-intuitive state that would not  be discovered without performing modal stability analysis on the quiescent, pure conduction base state. 
Because this eigenmode is characterized by two circulations of equal strength in the center of the valley, we refer to this eigenmode as the dual-circulation eigenmode. This eigenmode also satisfies a symmetry of reflection about the $y$ axis, which can be defined by the transformation:
\begin{equation}
        \mathcal{C}_2 : \left[u, v, b, p \right] \left(x, y \right) \mapsto \left[-u, v, b, p \right] \left(-x, y \right).
\end{equation}
The dual-circulation eigenmode is invariant under the transformation $\mathcal{C}_2$, and thus we say it is symmetric with respect to $\mathcal{C}_2$, or $\mathcal{C}_2$-symmetric in short. It can be shown that both the linearized and full N-S equations and boundary conditions are invariant under the $\mathcal{C}_2$ transformation. 
In other words, for any flow state in the valley $\mathbf{u}$, the transformed flow $\mathcal{C}_2(\mathbf{u})$ is also a solution. We also note that the action of this transformation creates a mirror image of the original flow state reflected about the vertical axis. Thus the symmetry of the dual-circulation eigenmode is only possible due to the symmetry of our valley geometry.

Beside the strongest circulation cells, each eigenmode also exhibits a series of progressively smaller eddies towards each corner. \citeauthor{moffatt1964viscous} \cite{moffatt1964viscous} studied the case of a sequence of eddies in a sharp corner due to a purely hydrodynamic (i.e. no thermal effects) motion in the fluid far from the corner with the Stokes equation, and predicted that an infinite sequence of eddies occurs in the corner with quickly diminishing strength. 
From an inspection of the eigenmodes, we can distinguish five distinct eddies in the central-circulation state, which can be observed due to the line integral convolution in Figure \ref{fig:eigenmodes}(a), and six eddies in the dual-circulation state, the smallest two being too weak to be visualized in the corners in Figure \ref{fig:eigenmodes}(c). Additional eddies can likely be identified with increased resolution in the corners, but we assume that any additional eddies are small enough in strength and size to have no effect on the overall flow. The progression of eddies we observe do not decrease in strength as strongly as predicted by \citeauthor{moffatt1964viscous}, and the ratio of the maximum $u$ velocity along the top boundary varies between the main circulation to the secondary circulations, and the secondary circulations to the corner eddies. The drop-off in strength between the main circulation and the secondary circulations is approximately 5 for the central-circulation case and approximately 12 for the dual-circulation case, whereas the drop-off between the secondary circulations and the corner eddies is approximately 100 for the central-circulation case and 600 for the dual-circulation case. The differences from the theory of \citeauthor{moffatt1964viscous} can be explained by the heated sloping walls in our case, which provides buoyancy force up the wall towards the corner along the entire wall, while \citeauthor{moffatt1964viscous}'s results are derived from a purely hydrodynamic motion of the fluid far from the corner. 

The reason for the onset of the central and dual-circulation instabilities can be explained from consideration of the dimensionless $\Pi_s$ parameter. For very small $\Pi_s$ values, the zero flow state can remain stable due to the conduction of the heat through the fluid, as well as the stabilizing effect of the background stratification, but as the surface heating increases, the stabilizing effect of the stratification is overcome, and convection begins to dissipate the additional heat. 

\begin{figure*}
    \centering
    \includegraphics[width=0.48\textwidth]{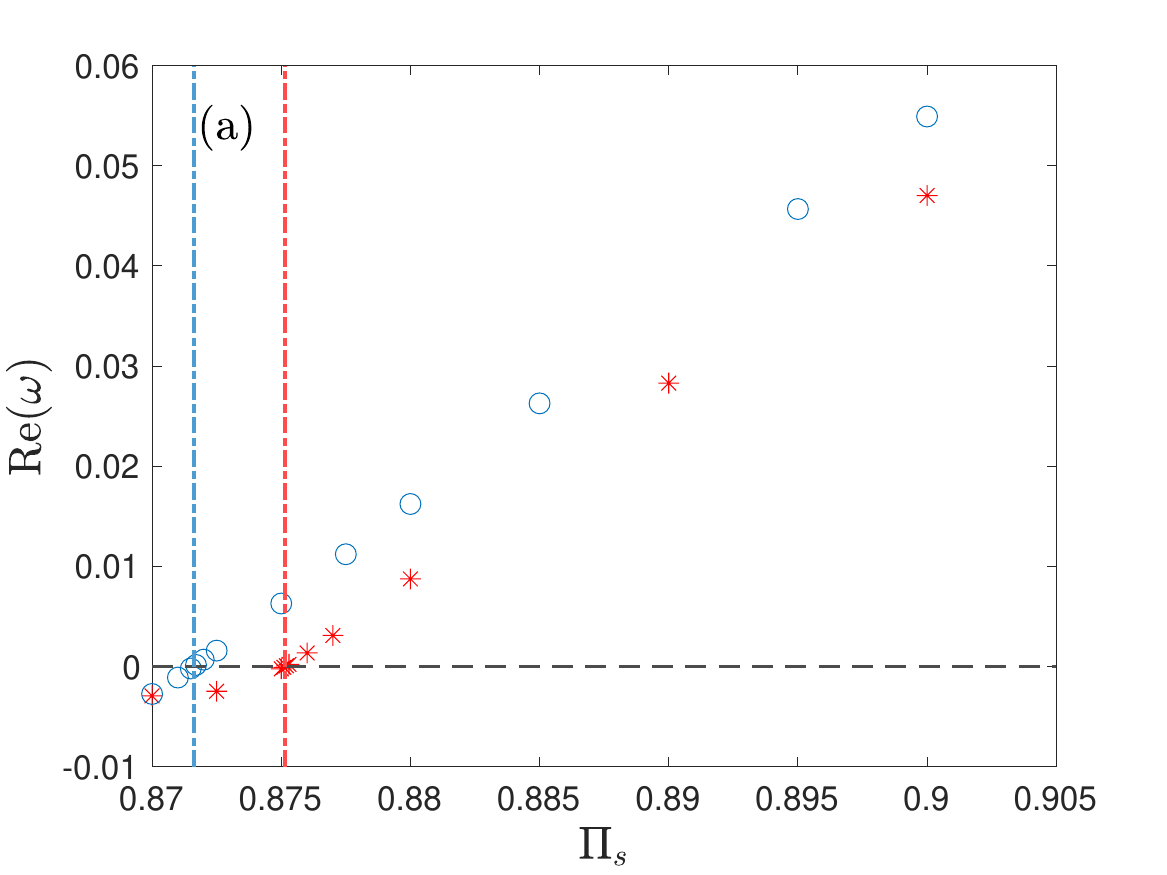}
    \includegraphics[width=0.48\textwidth]{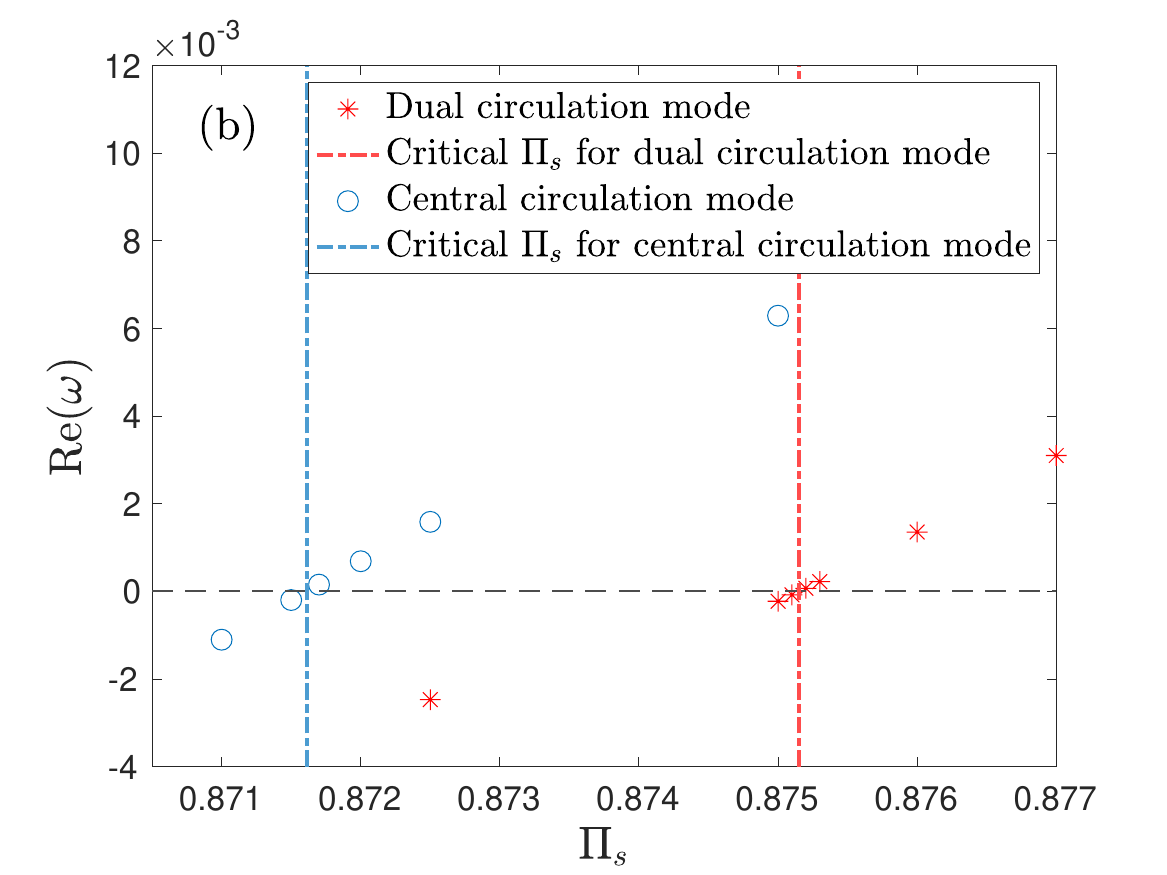}\hfill
    \caption{(a) Growth rate of dual-circulation  and central-circulation instabilities versus $\Pi_s$, with vertical lines denoting critical values. (b) Growth rate values zoomed into critical values.}
    \label{fig:growth_rate_vs_PIs}
\end{figure*}

The growth rate of each of the eigenmodes is plotted against $\Pi_s$ in Figure \ref{fig:growth_rate_vs_PIs}. 
We find that the critical value of the dual-circulation eigenmode to be approximately $0.87515$, whereas for the central-circulation eigenmode it is approximately $0.8716$. 
Thus, the central circulation has a lower critical value than the dual circulation, after which the growth rate of both modes appear to grow at approximately the same rate.
This indicates that the central-circulation eigenmode is the most unstable mode in a perfectly symmetric external configuration for all $\Pi_s$.

\begin{figure*}
    \centering
    \includegraphics[width=0.98\textwidth]{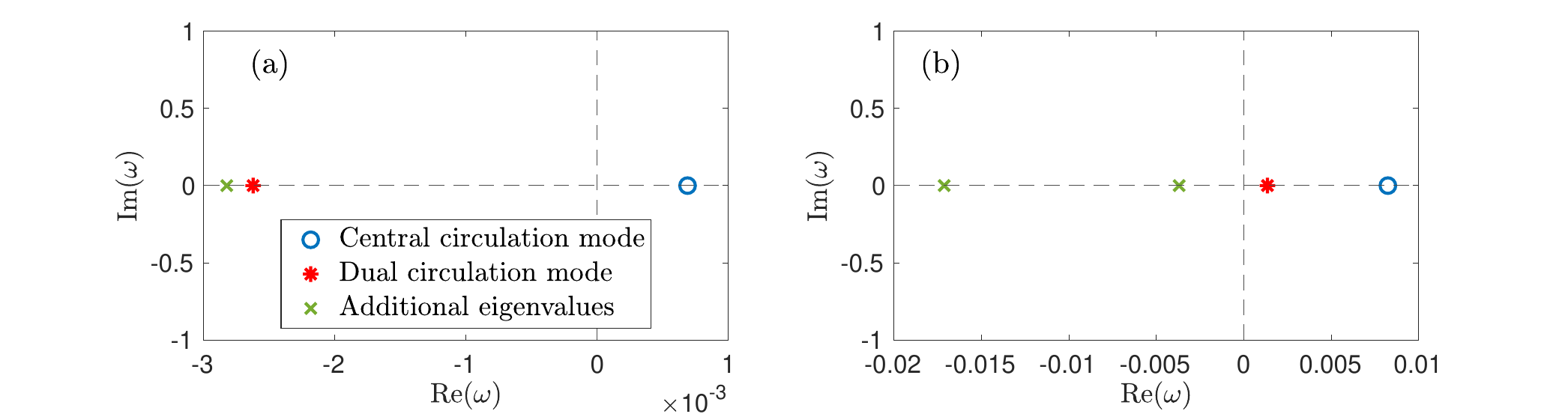}
    \caption{Eigenvalue spectrum for (a) $\Pi_s = 0.872$, and (b) $\Pi_s = 0.876$, showing the leading eigenvalues for the central-circulation and dual-circulation modes. Additional eigenvalues remain stable for all $\Pi_s$ investigated.}
    \label{fig:EV_spectra}
\end{figure*}

Figure \ref{fig:EV_spectra} shows the eigenvalue spectra for two $\Pi_s$ values near the critical values. We only show eigenvalues converged to a residual of $\epsilon < 10^{-6}$, which leaves only a few eigenvalues at each $\Pi_s$ value. We also specifically mark the eigenvalues that represent the central circulation and dual circulation eigenmodes, respectively. At $\Pi_s = 0.872$, in Figure \ref{fig:EV_spectra}(a), only the central circulation mode is unstable, having a growth rate $\mathrm{Re}(\omega) > 0$. As $\Pi_s$ is increased to 0.876 in Figure \ref{fig:EV_spectra}(b), the dual circulation growth rate increases and becomes positive. No other positive eigenvalues are found, and all eigenvalues have zero imaginary part.

\subsection{Steady-state Navier-Stokes solutions }

Next, we perform time-integration of the 3D Navier-Stokes equations, equations (\ref{eq:cont}-\ref{eq:buoy}), to obtain the steady-state solutions arising from the central and dual-circulation instabilities. Initial conditions for each simulation were defined by the analytical base flow, equation (\ref{eq:exact_buoy}), plus a small multiple of the eigenvector for each of the unstable eigenmodes. In this way, we can observe the initial exponential growth of the disturbance, and compare it to the growth rate predicted by LSA. 
It should be noted that while we perform 3D simulations with $L_z = 2$, all eigenmodes are 2D, which lead to only 2D steady states. 
Additionally, we run a simulation of $\Pi_s = 0.9$ with $L_z = 16$ with an initial random 3D disturbance, which produces a steady 2D state that is identical to our results with a 2D initial disturbance field. Thus, we can be confident that our assumption of 2D disturbances is not spurious due to lack of consideration of the third direction. The 2D circulation patterns in our 3D simulations agree qualitatively with the experimental smoke visualizations of \citet{holtzman2000laminar} who investigated convection in attic-shaped cavities with finite depth.

\begin{figure*}
    \centering
    \begin{subfigure}{0.49\textwidth}
        \includegraphics[width=\textwidth]{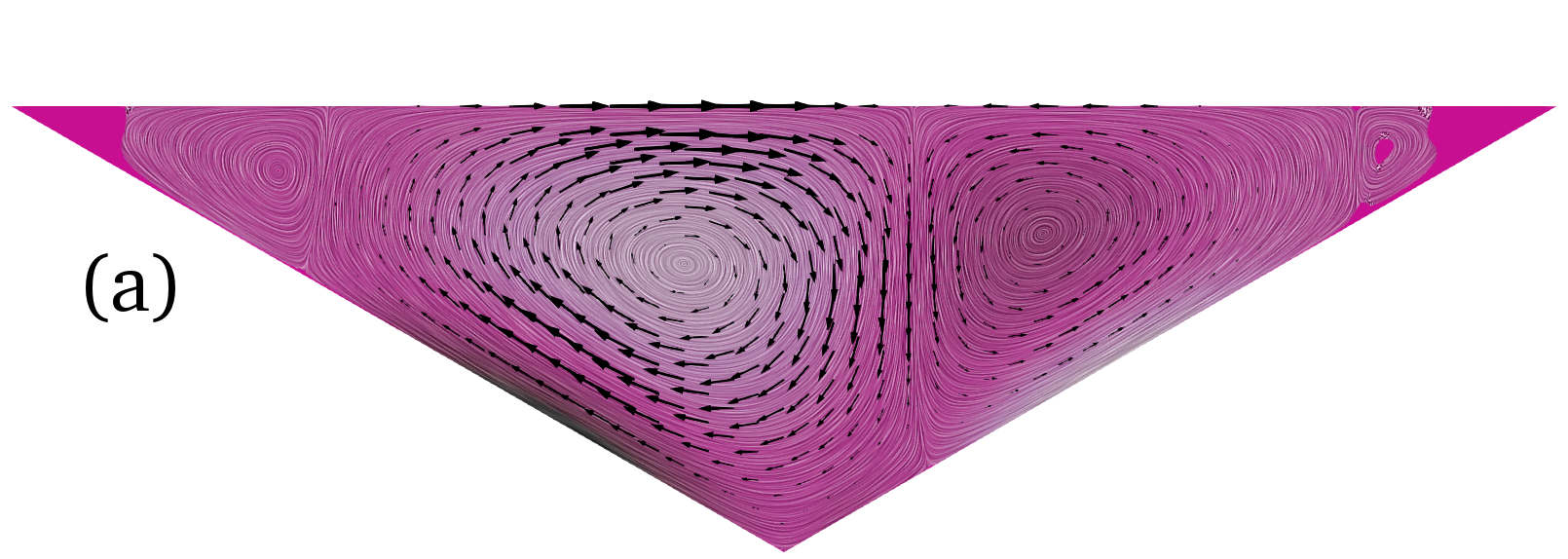} 
    \end{subfigure}
    \hfill
    \begin{subfigure}{0.49\textwidth}
        \includegraphics[width=\textwidth]{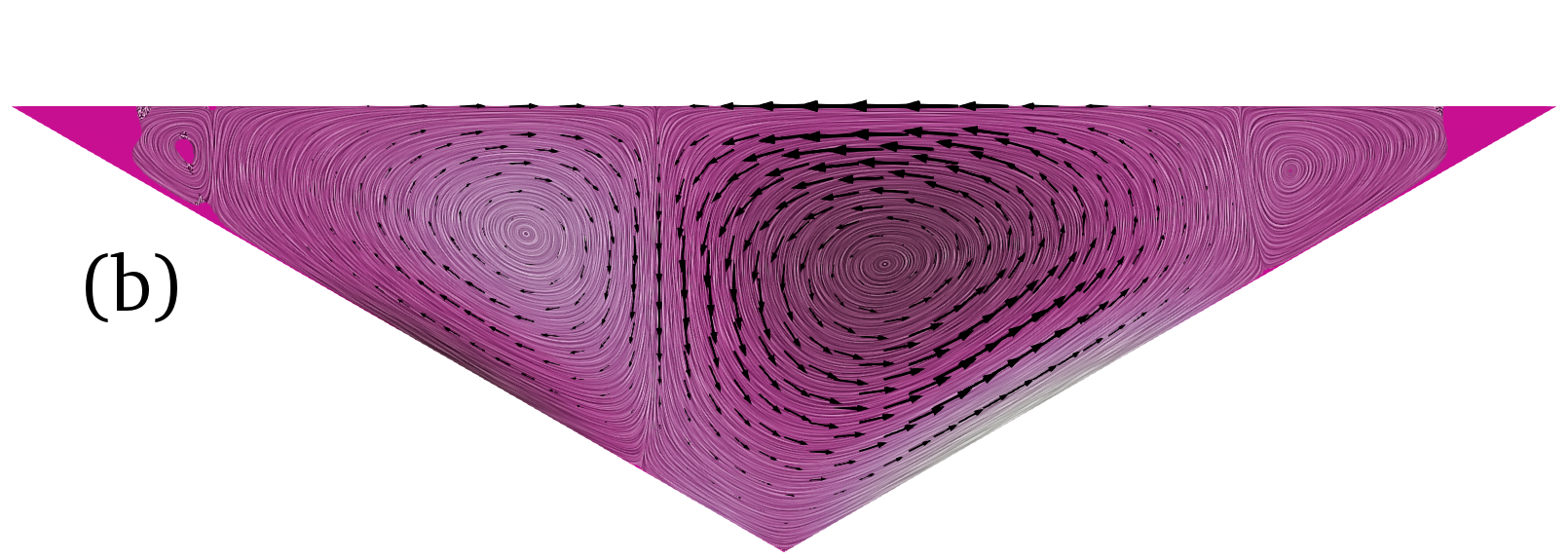}
    \end{subfigure}
    \hfill
    \begin{subfigure}{0.75\textwidth}
        \includegraphics[width=\textwidth]{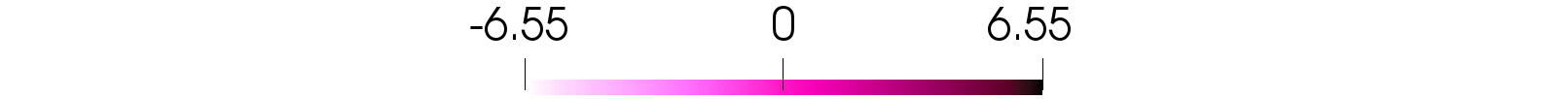}
    \end{subfigure}
        \hfill
    \begin{subfigure}{0.49\textwidth}
        \includegraphics[width=\textwidth]{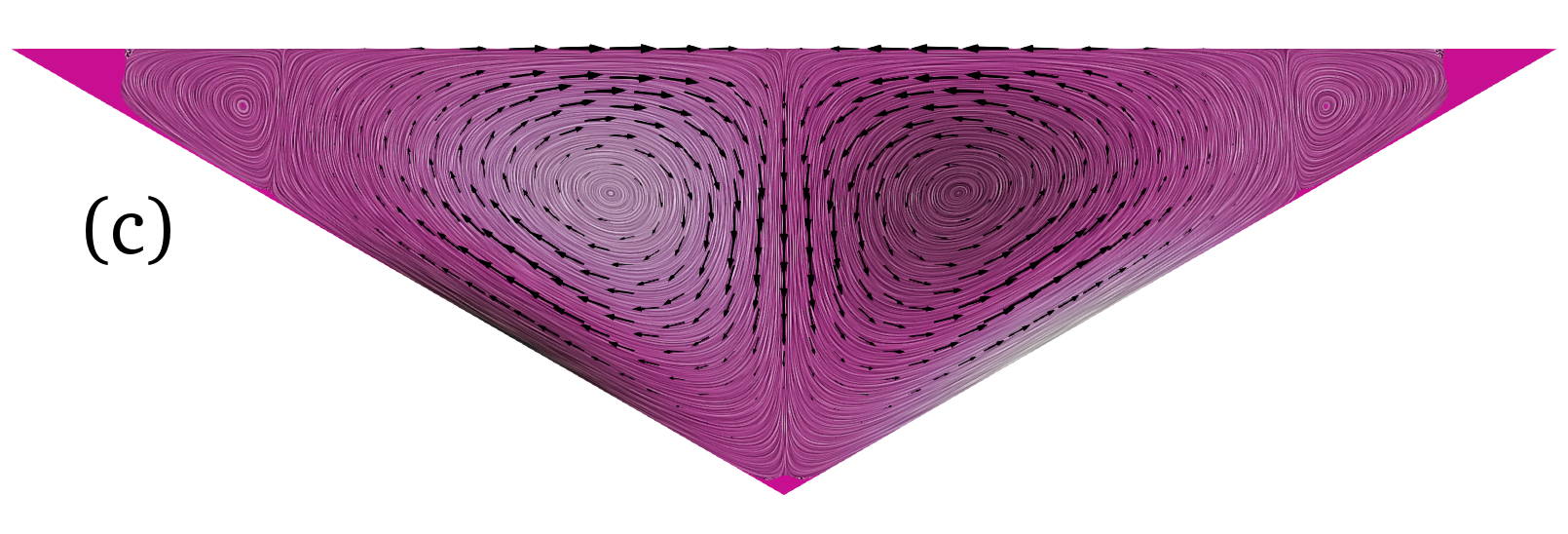}
    \end{subfigure}
        \hfill
    \begin{subfigure}{0.49\textwidth}
        \includegraphics[width=\textwidth]{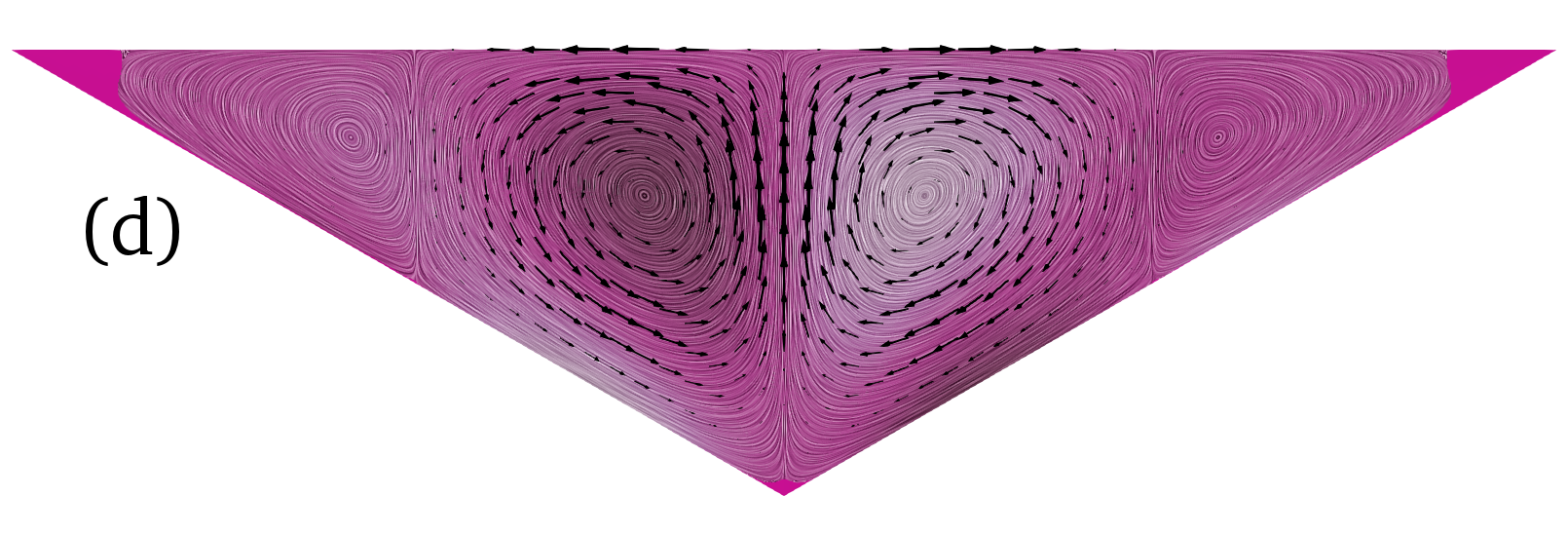}
    \end{subfigure}
    \caption{Visualization of vorticity along with velocity vectors of four possible steady-state solutions resulting from the central-circulation and dual-circulation eigenmodes at $\Pi_s = 0.9$: (a) Central circulation, clockwise, (b) central circulation, counterclockwise, (c) dual circulation, upslope, (d) dual circulation, downslope.}
    \label{fig:SS_profiles}
\end{figure*}

From the eigenvectors shown in Figure \ref{fig:eigenmodes}, we observe that the central-circulation eigenmode shows a clockwise main circulation, and the dual-circulation eigenmode shows two main circulations that travel down the sloping valley walls. These two eigenmodes represent four possible steady-state velocity profiles: the central-circulation state being either clockwise or counterclockwise, and the dual-circulation state being either upslope or downslope. Through manipulations of the initial eigenvector disturbance to our simulations, we can obtain steady-state profiles for each of these four states, as shown in Figure \ref{fig:SS_profiles}.

Focusing first on the central-circulation states shown in Figures \ref{fig:SS_profiles}(a) and \ref{fig:SS_profiles}(b), we can see that each is an exact reflection of the other about the $y$ axis, which can be explained by the symmetry of the valley geometry and the symmetry of the N-S equations about the $y$ axis. 
The steady-state profiles of the central-circulation state vary significantly from the corresponding eigenvector, shown in Figure \ref{fig:eigenmodes}(a), as these steady-state profiles are no longer symmetric about the $y$ axis. 
This is due to the fact, as stated earlier, that the transformation which defines the symmetry of the central-circulation eigenmode $\mathcal{C}_1$ is not invariant under the full N-S equations, and thus the $\mathcal{C}_1$-symmetry of the eigenmode must be broken in the nonlinear evolution. 

The nonlinear evolution leaves a new asymmetric steady-state profile, asymmetric in that each individual state, Figures \ref{fig:SS_profiles}(a) and \ref{fig:SS_profiles}(b), satisfies no symmetry about the $y$ axis with itself. 
Thus, the central-circulation eigenmode leads to symmetry breaking in the valley when considering the full N-S equations. Specifically this state breaks the $\mathcal{C}_2$ symmetry, which represents reflection about the $y$-axis and is the only symmetry that satisfies the nonlinear equations and geometry of the valley.
In fact, we see that while each steady-state solution is asymmetric, applying the $\mathcal{C}_2$ transformation to the state with the clockwise central circulation yields the other steady state with the counterclockwise circulation exactly, and vice versa.

The two steady states arising from the dual-circulation eigenmode are shown in Figures \ref{fig:SS_profiles}(c) and \ref{fig:SS_profiles}(d), and consist of two equal and opposite circulations in the center of the valley traveling up the sloping walls in one configuration, and down the walls in the other configuration.
In contrast to the central-circulation eigenmode, each individual steady-state profile of the dual-circulation eigenmode remains invariant under the $\mathcal{C}_2$ transformation. 
Thus we refer to the two dual-circulation states as the symmetric steady states. 
However, unlike the asymmetric states which can be defined as transformations of one another, the upslope and downslope symmetric states each represent a distinct state. In other words, while each \textit{individual} state is symmetric to itself under the $\mathcal{C}_2$ transformation, there is no symmetry relation \textit{between} the upslope and downslope states. This is the opposite of the scenario of the asymmetric, central-circulation steady states, in which each individual state satisfies no symmetry relation with itself, but there is a symmetry relation between the pair, namely the $\mathcal{C}_2$ transformation.
The lack of a symmetry relation between the dual-circulation states is due to the fact that the valley-shaped  geometry does not permit any horizontal axis of symmetry, and thus the opposite directions of circulation must lead to distinct final states. While both states originate from the same eigenmode, the nonlinear evolution leads to this difference in the final states due to the lack of any possible axis of symmetry. It should also be noted that negative versions of each symmetric state are not solutions, as this reversal of the velocity components does not satisfy the N-S equations. Instead, if the flow was reversed and simulated, each would evolve to the distinct steady upslope or downslope state shown in Figures \ref{fig:SS_profiles}(c) and \ref{fig:SS_profiles}(d).

The existence of the downslope flow state is counter-intuitive in a V-shaped enclosure heated from below; with heated sloping walls, one would only expect to see an upslope flow, and this expectation has been assumed by all previous studies of flows in heated, triangular cavities. While in nature the upslope flow is expected, mathematically the downslope flow state exists as a solution and can be achieved as a steady state in simulations only through careful initial conditions, and for sufficiently low $\Pi_s$ values. We should emphasize that we are only able to discover the downslope steady state due to our consideration of the quiescent, pure conduction base flow in LSA, which has not been considered in previous studies of stratified flows in valley-shaped triangular cavities. We have also performed simulations that confirm the existence of the downslope flow state at a lower slope angle of $10^{\circ}$. At larger slope angles, the downslope flow state is expected to become more unstable due to the greater component of the buoyancy force acting up the slope, and thus the downslope steady state may be unattainable in numerical simulations for such slope angles.

\begin{figure*}
    \centering
    \begin{subfigure}{0.49\textwidth}
        \includegraphics[width=\textwidth]{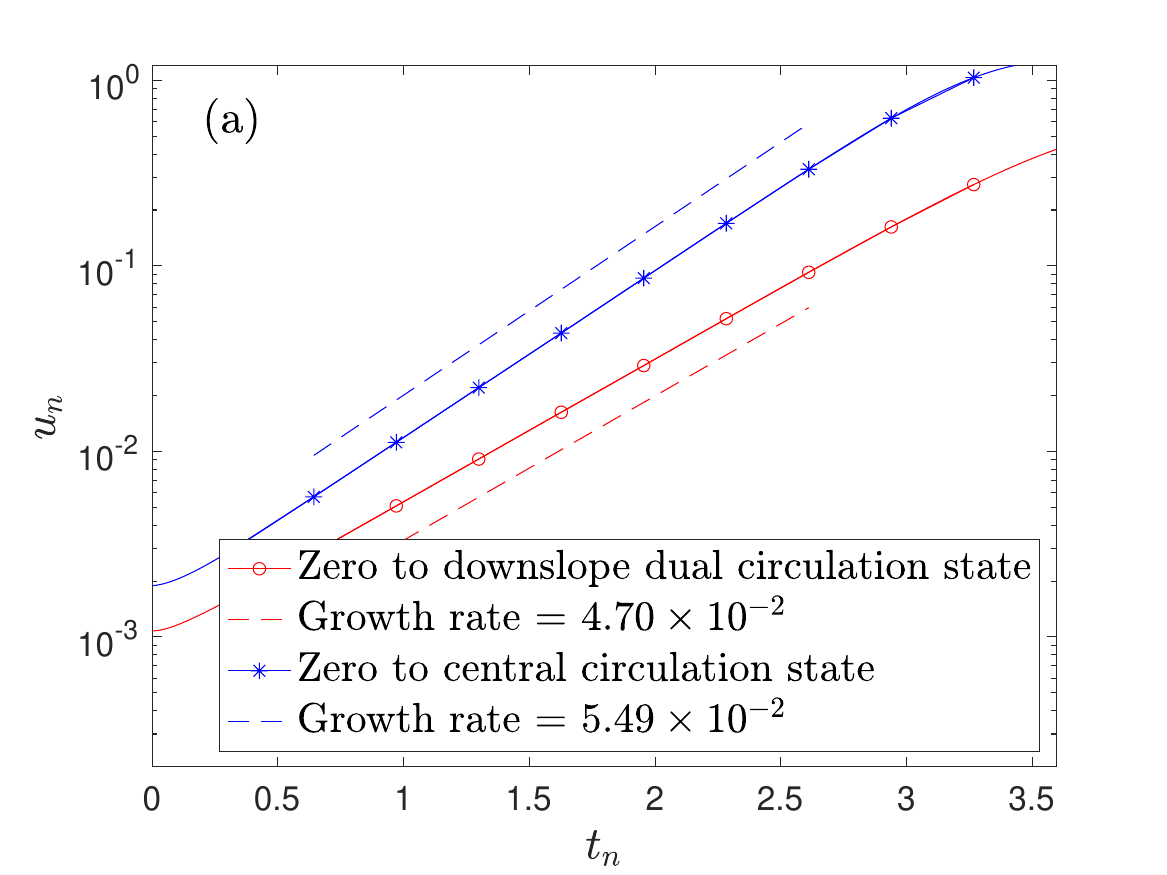}
    \end{subfigure}
    \hfill
    \begin{subfigure}{0.49\textwidth}
        \includegraphics[width=\textwidth]{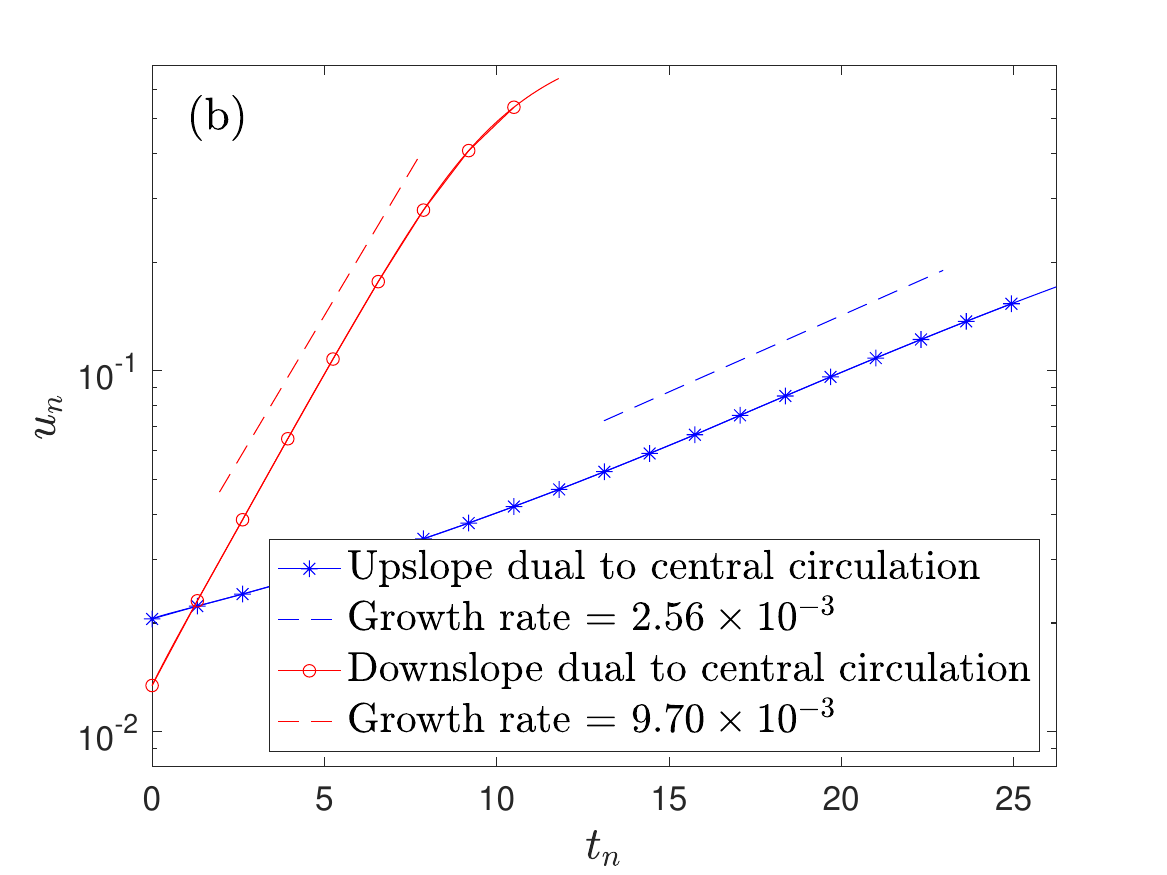}
    \end{subfigure}
    \caption{Comparison of initial growth of normalized $u$ velocity of simulation to growth rate predicted by LSA for (a) Zero state to the dual-circulation state at point $(0.36, 0.36, 0)$, and zero state to the central-circulation state at point $(0, 0.9, 0)$, and (b) Upslope dual-circulation state to central-circulation state, and downslope dual to central-circulation state both at point $(0, 0.9, 0)$. All cases at $\Pi_s = 0.9$. }
    \label{fig:GrowthRateVal}
\end{figure*}

The initial exponential growth of the central-circulation and the downslope dual-circulation disturbances in the nonlinear simulations is compared to the growth rate predicted by LSA in Figure \ref{fig:GrowthRateVal}(a). The initial conditions of both states consist of the analytical buoyancy and pressure base flow plus a small multiple of the eigenvector for the corresponding instability. We observe that the growth rates from the simulations match closely with the growth rates predicted from LSA. Additionally, the difference in slope between the two lines confirms the larger growth rate of the central-circulation instability.

Using the steady-state profiles obtained for the central-circulation and dual-circulation states, we now perform secondary linear stability analysis. First, our analysis shows that the asymmetric, central-circulation steady-state profiles are linearly stable at all $\Pi_s$ investigated here. However, LSA with the dual-circulation base flow results in an unstable eigenmode with an eigenvector similar to the primary central-circulation eigenmode shown in Figure \ref{fig:eigenmodes}(a). This was found to be true for all dual-circulation steady states. When this central-circulation eigenvector is added to the dual-circulation steady state in N-S simulations, the flow transitions from the symmetric, dual-circulation state to the same asymmetric, central-circulation state as seen previously. 
The mechanism for this transition is the slight increase in strength of one of the two central circulations, which will then gain in strength and move towards the center of the valley until a steady state identical to Figures \ref{fig:SS_profiles}(a) or \ref{fig:SS_profiles}(b) are reached.

The initial exponential evolution is shown in Figure \ref{fig:GrowthRateVal}(b) for both the upslope and downslope symmetric cases, and is compared to the growth rates predicted by LSA. Because the upslope and downslope symmetric states are distinct states, we observe different growth rates for each of the secondary instabilities, with the downslope state exhibiting a much larger growth rate than the upslope state. 
This makes sense since the state of downslope flow in a valley heated from the base is counter-intuitive and can only be achieved by consideration of a zero base state as well as careful initial conditions in simulations. The upslope state on the other hand, while also inherently unstable under the current conditions, represents a more natural state for flow in a heated triangular cavity and has been observed in previous numerical and experimental studies \cite{holtzman2000laminar,bhowmick2018natural}.
Compared to the primary instabilities, both secondary instabilities exhibit smaller growth rates, as can be seen from a comparison of the growth rates shown between Figures \ref{fig:GrowthRateVal}(a) and \ref{fig:GrowthRateVal}(b).

Comparing the possible transition scenarios from the zero-velocity base flow to the asymmetric steady state, it is worthwhile to point out a subtle yet significant distinction. 
In the first scenario when the original quiescent flow  is perturbed by the central-circulation eigenmode (Figure \ref{fig:eigenmodes}(a-b)), the reflection symmetry of the valley, or the $\mathcal{C}_2$-symmetry, is broken immediately due to the central-circulation eigenmode not satisfying this symmetry, and which evolves to the asymmetric steady state immediately from the eigenvector perturbation.
In the second scenario,  the flow is perturbed by the dual-circulation eigenmode (Figure \ref{fig:eigenmodes}(c-d))  and begins to evolve toward the symmetric steady state without losing the reflection symmetry of the initial perturbation in the process, until asymmetric perturbations begin to grow and only then is the symmetry of this steady state broken by its dominant instability mode that takes it to the asymmetric steady state. This describes the two possible paths by which the symmetry of the flow in the valley can be broken.

\subsection{Bifurcation diagram}

\begin{figure*}
    \centering
    \begin{subfigure}{0.49\textwidth}
        \includegraphics[width=\textwidth]{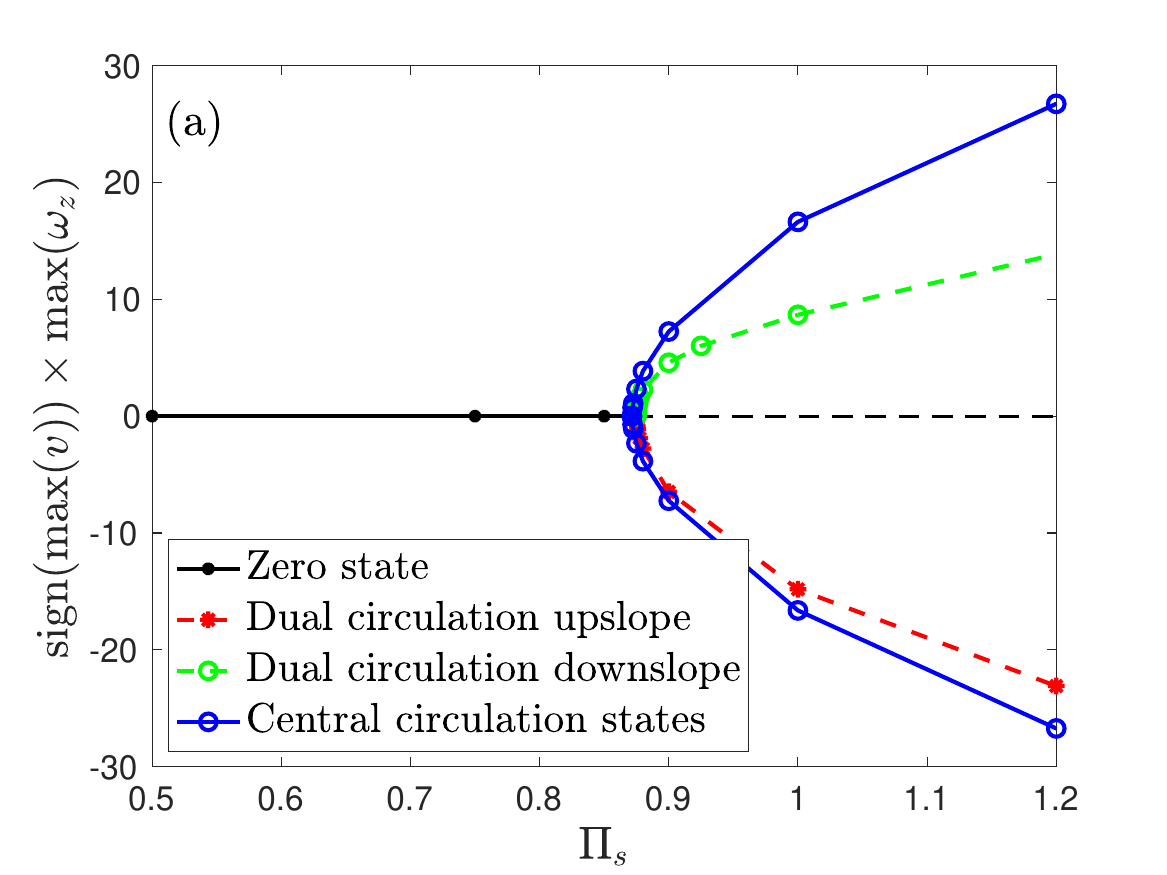}
        \label{fig:bifurcation_full}
    \end{subfigure}
    \hfill
    \begin{subfigure}{0.49\textwidth}
        \includegraphics[width=\textwidth]{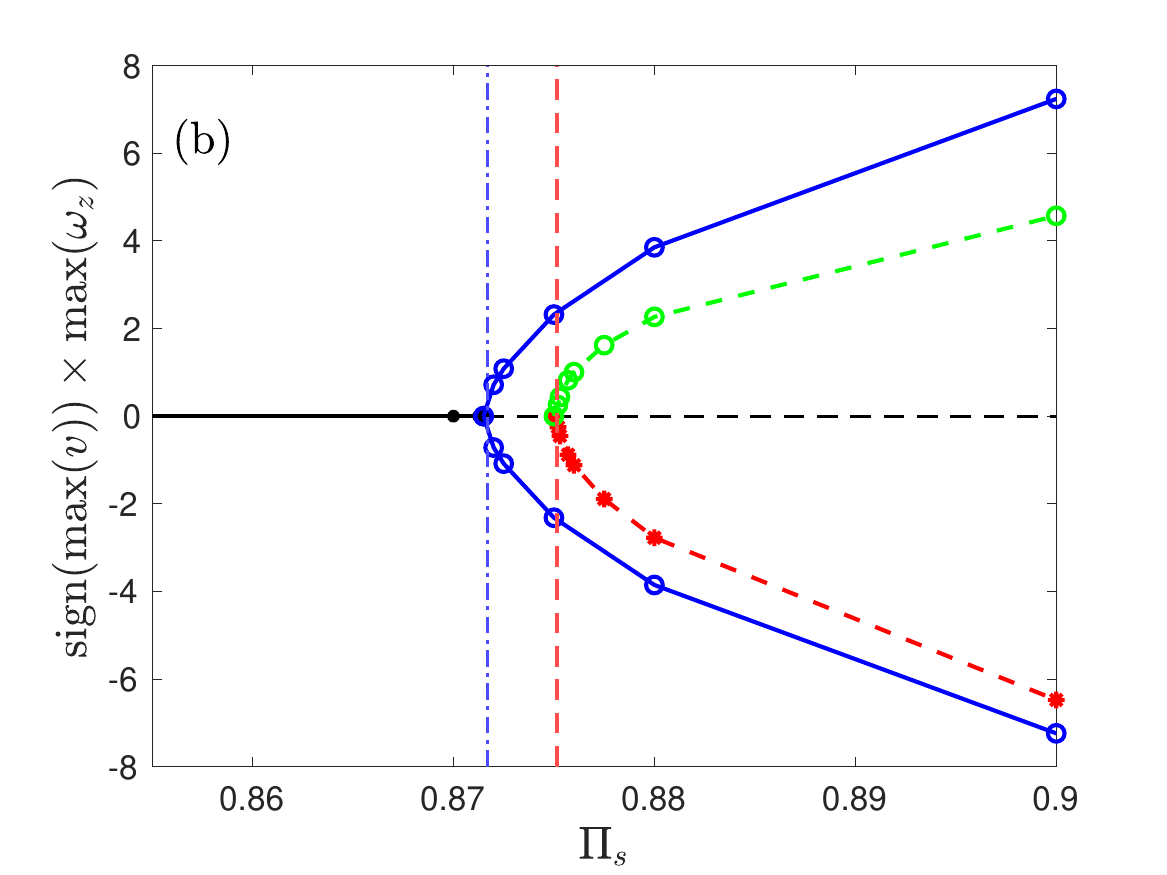}
        \label{fig:bifurcation_detail}
    \end{subfigure}
    \caption{Bifurcation diagram for increasing $\Pi_s$. The quantity on the $y$-axis is the maximum normalized vorticity $\omega_z$ in the $z$ direction obtained from 3D N-S simulations, multiplied by the sign of the maximum vertical velocity. (a) shows the transition from the zero state to the asymmetric, central-circulation states and the symmetric, dual-circulation states. Note the asymmetry of the red and green curves with respect to $\omega_z = 0$, (b) shows the same bifurcation plot zoomed into the critical value, along with critical values predicted by LSA shown as vertical lines.}
    \label{fig:bifurcation_figure}
\end{figure*}

From our simulations, we now characterize the change in the possible steady-state profiles with a change in our $\Pi_s$ parameter. We observe two primary instabilities, each of which lead to two possible steady-state profiles, and we draw a bifurcation diagram, shown in Figure \ref{fig:bifurcation_figure}, where we plot the absolute maximum vorticity (positive or negative) in the $z$ direction multiplied by the sign of the maximum vertical velocity $v$ as a function of $\Pi_s$. 
The choice of maximum vorticity is based on the two asymmetric, central-circulation states in which the maximum vorticity is equal and opposite in sign in each case due to the dominant central circulation and the opposite direction of flow. 
However, this is not true of the upslope and downslope symmetric states. Both of these individual states consist of dual circulations in the valley with equal and opposite vorticity values. To resolve this in a way that illustrates the bifurcation, we take the maximum positive vorticity value for each state and we multiply by the sign of the maximum vertical velocity $v$, which for the symmetric states occurs at the center of the valley and is opposite in sign. This creates the branching of the different states in the bifurcation diagram, with the negative branch representing the upslope steady state and the positive branch representing the downslope steady state. Because the two asymmetric states have exactly the same maximum vertical velocity value, this does not affect the outer branches at all. 

In the bifurcation diagram, at low $\Pi_s$ values, the zero velocity state is linearly stable and is represented by the solid black line. At the first critical $\Pi_s$ value, estimated to be approximately 0.8715, a bifurcation occurs leading the clockwise and counterclockwise asymmetric states. Because each of these states breaks the only symmetry of the valley geometry, namely the reflection symmetry about the vertical $y$-axis defined by transformation $\mathcal{C}_2$, this is a pitchfork bifurcation. Both of the branches are linearly stable, and are thus represented by the solid blue lines. 
Because the clockwise and counterclockwise central-circulation states are mirror images of one another, each has equal and opposite maximum vorticity, and thus the branches of this pitchfork bifurcation are perfectly symmetric with respect to the $\omega_z = 0$ axis, which is the form of a perfect pitchfork bifurcation.

At slightly larger $\Pi_s$, an additional bifurcation occurs from the unstable zero state, leading to the two possible dual-circulation states. The estimated critical $\Pi_s$ for this bifurcation is approximately 0.8751. Because the dual-circulation states are unstable to a secondary central-circulation eigenmode, these branches are represented by dashed lines in the bifurcation diagram. 
The steady-state profiles for each of these unstable states was obtained through simulating only half of the valley geometry with symmetric boundary conditions at the center. In this configuration, the symmetric steady states are stable, so we are able to accurately time march to the solution. We confirmed that this method provided the same symmetry solutions obtained through simulations of the whole valley. 

We observe a distinction in the inner bifurcation, which depicts symmetric solutions of upslope and downslope which are distinct from one another, but originate from the same eigenmode and the same base state. 
While the branches of the bifurcation bear some resemblance to a pitchfork bifurcation, a pitchfork bifurcation must break a specific symmetry within a system \cite{kuznetsov2004elements}. 
For instance, in our scenario, the two asymmetric states break the reflection symmetry of the valley about the vertical $y$-axis, and the two resulting states are symmetry-conjugated. However, when comparing the upslope and downslope states, no symmetry can be broken since the geometry lacks any possible horizontal symmetry axis, and because of this, the upslope and downslope state cannot be symmetry-conjugated, as in a pitchfork bifurcation. This is clear from the asymmetry in the branches of the inner bifurcation, as depicted in Figure \ref{fig:bifurcation_figure}. 

The asymmetry of the inner bifurcation diagram bears resemblance to an imperfect pitchfork bifurcation \cite{golubitsky2013singularities}, but this possibility is not supported by our DNS simulations, since both states emerge simultaneously from a single bifurcation point from the zero flow state. Further, because the zero-flow state must remain exactly zero for all $\Pi_s$, one branch cannot emerge continuously from the zero state, as observed in an imperfect pitchfork bifurcation. 

One might, alternatively, argue that this bifurcation structure arises from two separate bifurcations, given that the final upslope and downslope states are distinct. However, our analysis of the eigenvalue spectrum demonstrates that only two unstable modes exist: one corresponding to the asymmetric, central-circulation state and the other to both the upslope and downslope dual-circulation states. Furthermore, we can confidently assert that our results from LSA are accurate, and this structure is not influenced by any nonmodal or nonlinear effects due to our utilization of the zero-velocity base flow, in which case linear stability is equivalent to the nonlinear energy stability \cite{shir1968convective}. This also rules out any subcritical behavior below the critical point. We confirm this in our DNS simulations by starting from a steady state solution slightly above the critical point and lowering the $\Pi_s$ below the critical point, and the steady state is observed to decay back to the zero-velocity flow state. 

From these evidences, we can evaluate a number of possible bifurcation scenarios. First, the lack of separation at the critical point precludes the possibility of an imperfect pitchfork bifurcation, and the lack of subcritical behavior or the nonexistence of either symmetric state below the critical value precludes a transcritical bifurcation for either state.
Consequently, the evidence from both LSA and DNS indicates that two related fixed points emerge from the zero state at the same critical value, indicative of an unusual bifurcation. The distinctive feature of this bifurcation is that it gives rise to two related yet distinct states stemming from the same eigenmode without breaking the sole symmetry of the valley. 

\begin{figure*}
    \centering
    \begin{subfigure}{0.49\textwidth}
        \includegraphics[width=\textwidth]{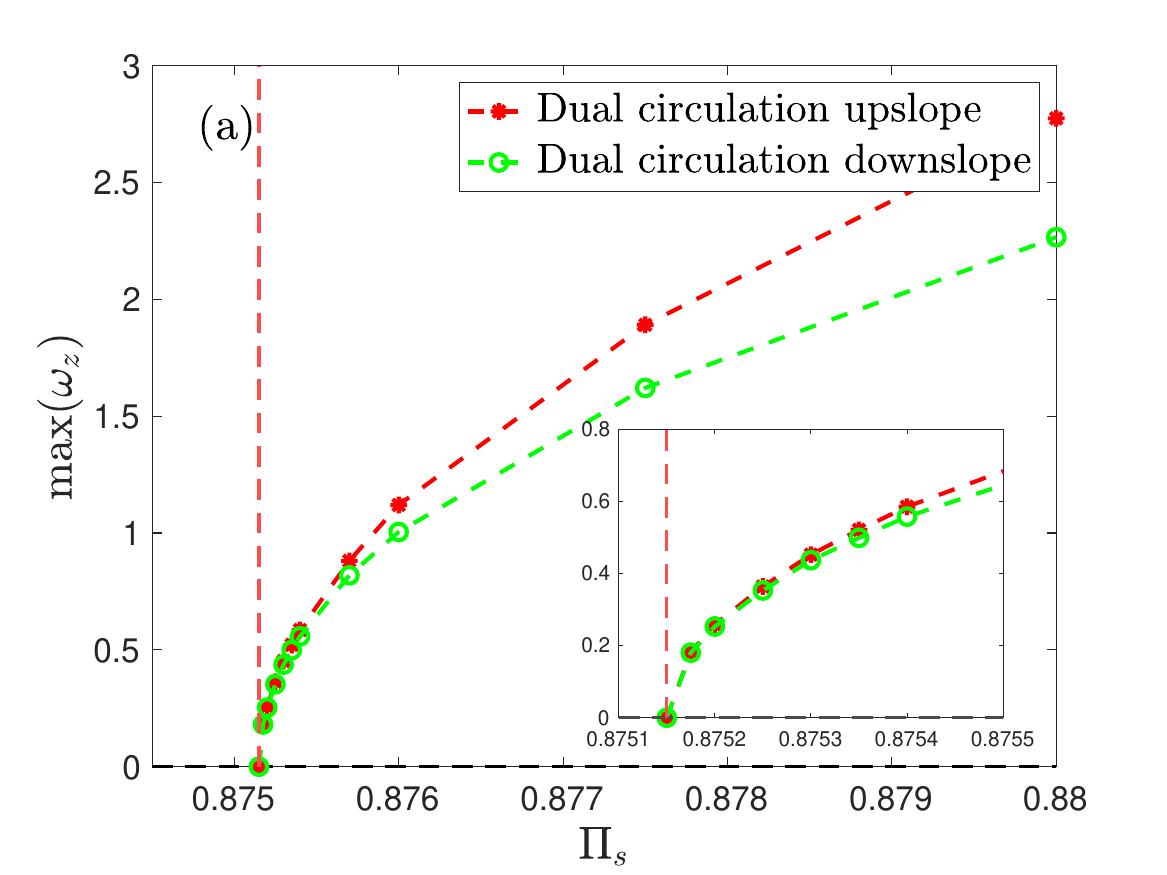}
        \label{fig:UpslopeVsDownslopeA}
    \end{subfigure}
    \hfill
    \begin{subfigure}{0.49\textwidth}
        \includegraphics[width=\textwidth]{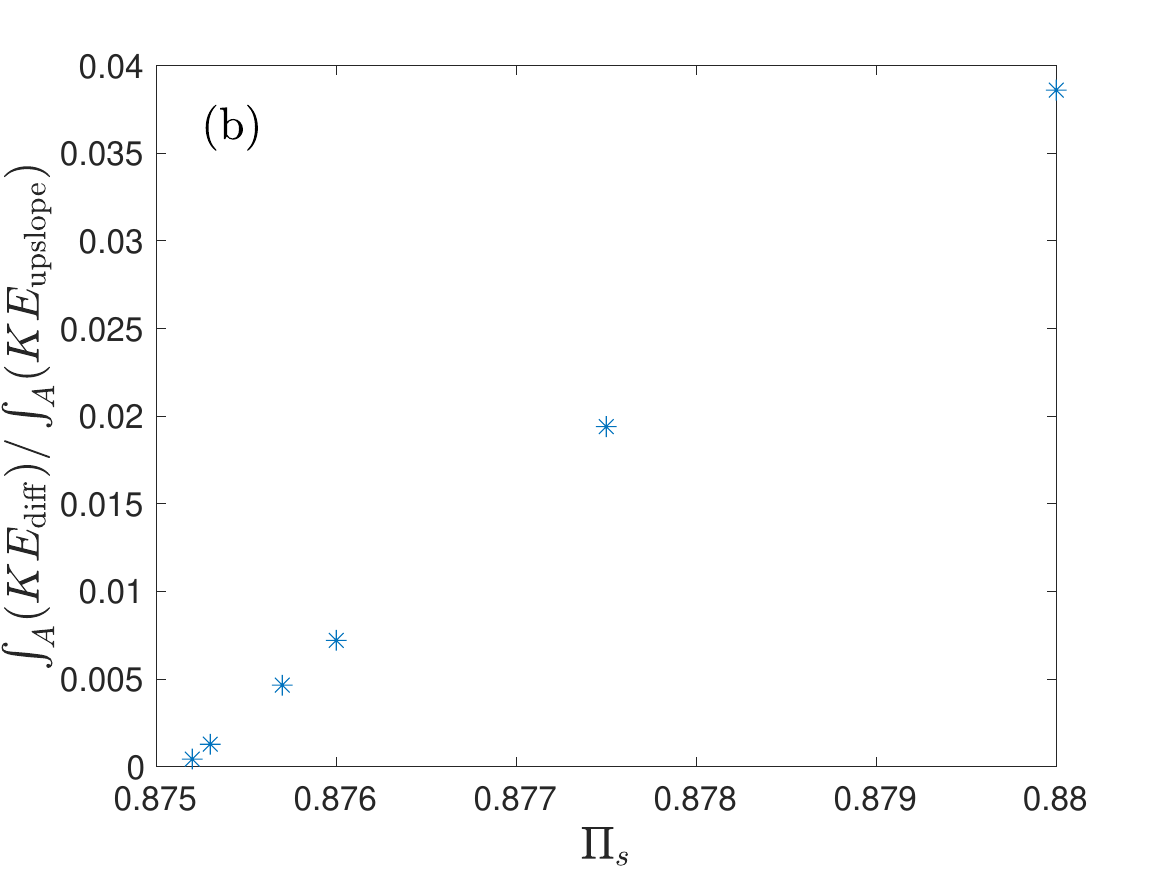}
        \label{fig:UpslopeVsDownslopeB}
    \end{subfigure}
    \hfill
    \begin{subfigure}{0.49\textwidth}
        \includegraphics[width=\textwidth]{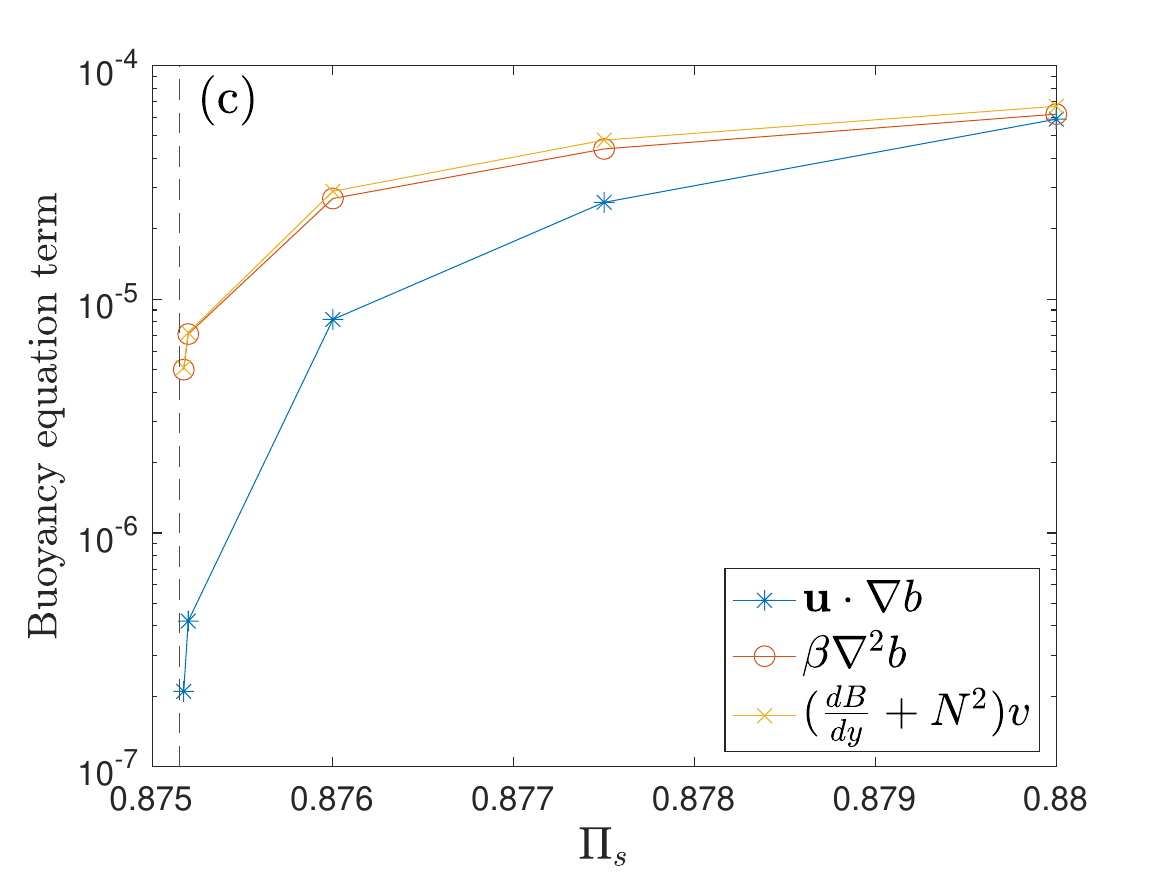}
        \label{fig:UpslopeVsDownslopeC}
    \end{subfigure}
    \caption{Comparison of upslope and downslope dual circulation steady states near the critical value. (a) Maximum $z$-vorticity versus $\Pi_s$, same values as in Figure \ref{fig:bifurcation_figure}(a) but with same sign. Inset shows the data further zoomed into critical value.} (b) Integral of difference KE between upslope and downslope states versus $\Pi_s$, normalized by integral of KE of upslope state. (c) Maximum value of each term in the buoyancy equation at the symmetric, upslope steady state versus $\Pi_s$ near the critical value.
    \label{fig:UpslopeVsDownslope}
\end{figure*}

We further examine the relation between the two states of this inner bifurcation in Figure \ref{fig:UpslopeVsDownslope}. 
Figure \ref{fig:UpslopeVsDownslope}(a) compares the maximum vorticity for the upslope and downslope states near the critical value, which shows that the difference between the states grows with $\Pi_s$, and that very near the critical value, the maximum vorticity of each state is close to identical. This indicates that close to the critical value the velocity profile of each state is very close to the reverse of each other. This makes sense, for when considering the linearized equations, the eigenvector corresponding to the dual-circulation upslope and downslope states respectively are scalar multiples of each other by $-1$, both of which satisfy the linearized Navier-Stokes equations. Of course, when considering the nonlinear evolution, this transformation does not satisfy the full Navier-Stokes equations, and thus the upslope and downslope states diverge in nonlinear simulations. However, very close to the critical value the nonlinear effects remain small and the steady-state solutions are close to scalar multiples of one another. This is shown in Figure \ref{fig:UpslopeVsDownslope}(b) which shows the kinetic energy (KE) difference between the upslope and downslope states versus $\Pi_s$. Specifically, this is calculated as the area integral of the difference kinetic energy, where the difference velocity is defined by $\mathbf{u}_{\mathrm{diff}} = \mathbf{u}_{\mathrm{upslope}} + \mathbf{u}_{\mathrm{downslope}}$, normalized by the integral of the kinetic energy of the upslope state. Thus, for upslope and downslope velocity profiles that are the exact reverse of one another, the difference velocity is zero. 
Figure \ref{fig:UpslopeVsDownslope}(b) shows that the difference remains low at very low $\Pi_s$ values. 
This indicates that very close to the critical value, the upslope and downslope velocity profiles are very close to the reverse of one another.
The reason for this is further confirmed in Figure \ref{fig:UpslopeVsDownslope}(c), which shows the maximum value of each term in the buoyancy equation for increasing $\Pi_s$. As expected, very close the the critical value, the nonlinear term is about two orders of magnitude smaller than the two linear terms. 
Therefore, very close to the critical value, the nonlinear term plays a small role and the steady state profiles are determined mainly by the linearized Navier Stokes equations. Because the linearized Navier Stokes equations admit solutions that are equal and opposite in sign, the upslope and downslope states are close to the reverse of one another near the critical value, which manifests as an increasing symmetry between the two branches of the bifurcation diagram.

\section{Conclusion}

We examined the instabilities of stably stratified laminar flows within a valley-shaped enclosure heated from below with unique boundary conditions and a representation of ambient stratification in the governing equations, following the Prandtl model for slope flows. We performed  linear stability analysis (LSA) as well as three-dimensional Navier-Stokes (N-S) simulations with 2D and 3D perturbations to validate our results. For a fixed slope angle and Prandtl number, the flow behavior is delineated by two dimensionless parameters: the stratification perturbation parameter $\Pi_s$, and $\Pi_h$, a new addition to the parameter space of stratified heated triangular cavities, and which is related to the buoyancy number employed in related studies \cite{yalim2018vertically}. In the results presented here, we focus only on the effect of the $\Pi_s$ parameter while keeping the remaining dimensionless parameters fixed. When $\Pi_s$ assumes very small values, indicative of subtle surface heating or very strong background stratification, a stable quiescent conduction state persists, for which we derive exact solutions for the pressure and buoyancy field. 

Above a critical $\Pi_s$ value, two main instabilities come to the forefront as determined by LSA. One eigenmode reveals a single dominant circulation at the valley's center, while the other displays two circulations of equivalent yet opposing intensity within the central valley region. The critical value for the central-circulation eigenmode is slightly lower than that for the dual-circulation eigenmode. 
We find that both eigenmodes are symmetric about the vertical axis, but each satisfies a different symmetry, with the symmetry of the central-circulation mode satisfying only the linearized equations, while the dual-circulation mode satisfies the reflection symmetry about the $y$-axis for the full nonlinear equations.

The dual-circulation eigenmode gives rise to two types of symmetrical steady-state profiles, where one features   circulations containing upslope flow, and the other features downslope flow. While these states maintain the same reflection symmetry as the eigenmode, their dissimilarity results from the absence of any feasible symmetry about the $x$-axis within the valley  geometry.
The second type of instability, the central-circulation eigenmode, gives rise to a pair of asymmetric steady-state profiles, each characterized by either clockwise or counterclockwise dominant circulations in the valley-shaped enclosure. While the symmetry of the eigenmode is broken by the nonlinear evolution to the steady-state profiles, the clockwise and counterclockwise asymmetric states are perfect reflections of one another about the vertical axis.

The secondary stability analysis reveals that the symmetric, dual-circulation steady-state profiles are prone to further instability, leading to the emergence of the steady state asymmetric circulation. 
This creates a bifurcation diagram with a unique bifurcation leading to the upslope and downslope symmetric states nested within a perfect pitchfork bifurcation, representing the steady, asymmetric circulation states.

The inner bifurcation is of special interest because it represents a bifurcation to two states, the upslope and downslope states, which maintains the reflection symmetry about the vertical axis, but are also distinct from one another due to the lack of any possible horizontal axis of symmetry, unlike an analogous case in a rectangular geometry. The difference between the upslope and downslope states can be seen in the asymmetry of the branches of the inner pitchfork bifurcation about the $\omega_z = 0$ axis in the bifurcation diagram. 
We have evaluated a number of possible bifurcation scenarios that may match this behavior. The lack of any symmetry breaking rule out the possibility of a pitchfork bifurcation, while the lack of separation at the critical point rules out an imperfect pitchfork bifurcation. To this end, the observed bifurcation scenario does not align with any canonical types of bifurcations \cite{kuznetsov2004elements}. Consequently, we defer the identification of this bifurcation as an open question. 

The present study adds a number of important observations about the essential flow dynamics in stratified heated valley-shaped enclosures that have not been previously recognized. First, given our chosen boundary conditions combined with the presence of an ambient stratification, we have identified the existence of a quiescent, pure conduction state which is stable with very low surface heating or very large stratification. With the discovery of this quiescent base state that has been overlooked in previous works, we are able to characterize the complete flow behavior within the parameter space through the use of LSA and nonlinear N-S simulations, which reveal five possible equilibrium states, as depicted at the larger $\Pi_s$ values in the bifurcation diagram. This characterization also reveals the existence of the counter-intuitive downslope, symmetric flow state, another previously unidentified flow state in heated, stratified valleys. While this downslope state is inherently unstable, our use of LSA along with careful initial conditions in 3D N-S simulations enables us to identify that this is an additional possible steady state solution.
Another important finding in the current study is the instability of the symmetric steady state and precedence of the asymmetric steady state over the symmetric state.
Prior studies of flows in heated, triangular cavities suggest that the symmetric, dual-circulation state is the base state which only transitions to the asymmetric state after a pitchfork bifurcation occurs \citep{ridouane2006formation,bhowmick2018natural}. In contrast, under the current configuration, we find the base state can be a quiescent, steady, pure conduction state. With this base state, the asymmetric state manifests in the valley-shaped enclosure as a result of the primary instability, with a larger growth rate and lower critical $\Pi_s$ value in comparison to the symmetric state. In addition, we find the symmetric state to be unstable at all parameter values considered here. This observation is important because, for a random perturbation to the quiescent base flow, we would only expect to obtain asymmetric steady-state profiles, and not the symmetric dual-circulation state. We confirmed this expectation through 3D Navier-Stokes simulations with random initial conditions for $L_z$ between 2 and 16 at $\Pi_s = 0.9$, all of which are observed to converge to the asymmetric steady state.

Our findings reveal that the combination of linear stability analysis and nonlinear flow simulations can be an effective and precise approach for analyzing the dynamics of steady laminar flows within a stably stratified, heated valley-shaped enclosure. Within this context, we have demonstrated that, for a given set of dimensionless parameters, the Navier-Stokes equations admit at least five unique solutions, including symmetric and asymmetric convection patterns and an additional quiescent, pure conduction state. However, the asymmetric state is stable at all parameter values investigated here whereas the symmetric state is unstable, suggesting a higher likelihood of observation of the asymmetric state in natural settings. This phenomenon is anticipated to be even more pronounced when considering the inherent asymmetry and heterogeneity present in real valleys.

\section*{Acknowledgments}
This material is based upon work supported by the National Science Foundation under Grant No. (1936445) and (2203610) and by University of Pittsburgh Center for Research Computing, RRID:SCR\_022735, through the resources provided. Specifically, this work used the H2P cluster, which is supported by NSF award number OAC-2117681.

\newpage
\bibliography{apssamp}

\end{document}